\documentclass[bm]{aa}
\input psfig.tex
\usepackage{graphicx}
\usepackage{natbib}

\usepackage{array}
\usepackage{graphics}
\usepackage{latexsym}
\usepackage{amssymb}
\usepackage{amsmath}
\usepackage{fancyhdr}
\usepackage{morefloats}
\usepackage{bm}
\bibpunct{(}{)}{;}{a}{}{,}

\begin{document}
\title{Tidal Effects on the Spatial Structure of the Local Group}

\author{Stefano Pasetto$^{1,2}$ \&  Cesare Chiosi$^{3,4}$}

\institute{$^1$ Astronomisches Rechen-Institut, Zentrum f\"ur Astronomie der Universit\"at Heidelberg,  M\"onchhofstr. 12-14,
D-69120, Heidelberg, Germany \\
$^2$ Max Planck Institut f\"ur Astronomie, K\"onigstuhl 17, D-69117, Heidelberg, Germany\\
$^3$  Department of Astronomy, Padova University, Vicolo
dell'Osservatorio 2, I-35122, Padova, Italy\\
$^4$ Max Planck Institut f\"ur Astrophysik, Karl-Schwarzschild-Str. 1, D-85748, Garching, Germany\\
 \email{{spasetto\char64ari.uni-heidelberg.de},{cesare.chiosi\char64unipd.it}}}
\date{Received: ;  Revised: ; Accepted...}

\titlerunning{Tidal effects on the Spatial Structure of the Local Group}
\authorrunning{S. Pasetto \&\ C. Chiosi}

\abstract{ \textsc{Aims}. The spatial distribution of galaxies in
the Local Group (LG) is the footprint of its formation mechanism and
the  gravitational interactions among its members and the
external massive galaxies or galaxy groups.
Recently, Pasetto \& Chiosi (2007), using a 3D-geometrical
description of the spatial distribution of all the members of the LG
(not only the satellites of the MW and M31) based on present-day data on
positions and distances, found that all galaxies (MW, M31, their
satellites, and even the most distant objects) are confined within a
 slab of about 200 kpc thickness. Examining  how external galaxies or
groups would gravitationally affect (and eventually alter) the
planar structure (and its temporal evolution) of the LG, they found
that the external force field acts parallel to the plane determined by
geometry and studied this with the Least Action Principle.

\textsc{Methods}. In this paper, we have thoroughly
investigated the role played by the tidal forces  exerted by external
galaxies or galaxy groups on the LG  galaxies
(the most distant dwarfs in particular)  in shaping their
large scale distribution. The idea based on the well known effect
of tidal interactions, according to which a system of mass-points
can undergo not only tidal stripping but also tidal compression
and thus become flatter.

\textsc{Results}.  Excluding the dwarf galaxies tightly bound to the MW
and M31,  the same tidal forces can account for the planar
distribution of the remaining dwarf galaxies. We analytically
recover the results of Pasetto \& Chiosi (2007) and prove that a
planar distribution of the LG dwarf galaxies is compatible with the
external force field. We also highlight the physical cause of this
result.
 }

\maketitle

\section{Introduction}\label{Introduction}

Over the years, many  attempts have been made to find the spatial
distribution of the LG  galaxies. Limiting ourselves to a few
pioneering  studies and a few contributions in the past decade,
 \citet{1959ApJ...130..705K} and
\citet{1989MNRAS.240..195R} suggested a planar distribution based on
studies of the LG dynamics; \citet{2000AJ....119.2248H} found a
flat ellipsoid with axial ratios $\left(a,b,c\right) = \left(1.00,
0.51, 0.19\right)$ which is not too different from a plane; finally
\citet{1979ApJ...228..718K}, \citet{ 1999IAUS..192..447G} and
\citet{1995AJ....110.1664F} suggested that the satellite dwarf
galaxies of the MW and M31 lie on planes \footnote{Closely
related to this problem is the issue of the
anisotropic distribution of inner sub-haloes with respect to larger
haloes  in relation to the Holmberg effect
\citep{1969ArA.....5..305H} with dissimilar results, e.g.
 \citep{2004MNRAS.348.1236S, 2006MNRAS.369.1293Y}. What matters here (and is still
debated) is whether disruptions and tidal effects can create the apparent
polar alignment of the dwarf satellites around the host galaxy
 or, for the particular case of the LG, the position
of the dwarf galaxies is the consequence of peculiar directions of
pre-existing cosmological filaments.}.

More recently, \citet{2005A&A...431..517K} and \citet{2007MNRAS.374.1125M} suggested a planar
distribution of the satellites of the MW, which however could also
be explained as a consequence of the distribution of sub-haloes
\citep{2005ApJ...629..219Z} in the early cosmological stages
\citep{2005A&A...437..383K}.  The same problem has been investigated
by \citet{2006AJ....131.1405K} for the satellites of M31 with similar conclusions.

Starting from the basic idea that an off-center
hydrodynamical collision occurred some 10 Gyr ago between the
primordial gas-rich M31 galaxy and the MW, and compressed the halo
gas to form all the LG dwarf galaxies, \citet{2005PASJ...57..429S}
suggested  that  the new-born dwarf galaxies would be located near the orbital
plane of the MW and M31. They argued that this view is also
sustained by the visual inspection of the 2D sky distribution of the
LG members and  that a well-defined plane of finite thickness is
found, within which most of the member galaxies are confined.

\citet[][ hereafter PC07]{PC07}
attacked the problem from a completely different perspective.
In summary, adopting known data on positions and distances and  making
use of analytical geometry,  they looked for the plane that
minimizes the distances of all galaxies in the LG to it (not only the MW
and M31 and their satellites, but also the distant dwarfs). The
second part of their study was to find a dynamical justification for
the planar distribution. To this aim, they applied the Hamilton
Method (Minimum Action) to investigate the dynamics of the LG complex
and the action of the gravitational forces exerted by external
nearby galaxies or groups. They found that the planar distribution
is fully compatible with the minimum action and that the external
force field is likely compatible with the plane. Such a field pulls the LG
galaxies  along,  without altering their planar distribution.
Special care was taken to evaluate the robustness of the result.

To somehow account for the different results obtained by
\citet{2005A&A...431..517K}, \citet{2006AJ....131.1405K}, and
\citet{2005PASJ...57..429S}, it is worth recalling here that the
various studies did not use the same galaxy sampling, start
from the same  working physical hypotheses nor deal with the same
dynamical regime.  In brief:

(i) The  planes for the MW and M31 satellites \citep[][
respectively]{2005A&A...431..517K,2006AJ....131.1405K} are   of a very
local nature as they are the consequence of   strong collisional
dynamics with the host galaxy (hereafter HG). No easy explanation can be found to
secure that these planes will survive for long time (more
than a few dynamical time scales) due to the peculiar proper motions
of the dwarfs that determine these planes: see, e.g., the ideas in \citet{1983IAUS..100...89L, 1995MNRAS.275..429L, 2002ApJ...564..736P}
applied then by \citet{2008ApJ...680..287M}. What would be role of
distant dwarfs in determining the structure of the whole LG is
simply not cosnidered. Therefore, these planes cannot be extended
to the whole LG.

(ii) In  \citet{2005PASJ...57..429S} the solution for a
common plane is based on the ad hoc  initial hypothesis concerning the
origin of the angular momentum.  The
sample of dwarf galaxies used to  determine  the plane is
limited to the  satellites of the two HGs. Finally, the
 orbits of these dwarfs are constrained to lay on this plane.

(iii) In PC07,  the common plane is chosen by assuming that it
contains the MW and M31 and minimizing the distances of all
remaining LG galaxies to this plane, including also the distant ones.  This can be
justified by considering that the HG satellites are strongly
influenced by local dynamics (with continuous modification of their
orbits, including possible captures by the HGs) and that if a common planar
distribution for all LG galaxies exists, this should  be brought
into evidence by the more external galaxies. They are  much
less likely to be affected by strong interactions with one of the HGs and
therefore more sensitive to the influence of external galaxies and/or
groups. The plane found by PC07 is actually a slab of about 200 kpc
thickness, i.e. it is worth mentioning that the largest apo-center of the
HG satellites most probably falls inside this slab.

In addition, PC07 have shown that the external force
field runs parallel to their plane. It is likely that among the
galaxies of the LG those that feel the external action the most are the
dwarfs not tightly bound to any HG. \textit{In  other words, this
group of dwarfs could act as a tracer of the external force field.
Our aims is to find a gravitational action that is able to induce, on a
long time scale, a sort of extended slab. Tidal
forces are known to engender this kind of response.}

The main goal of this study is to highlight the physical
nature of the results obtained by PC07  that had a rather complicated
dynamical description requiring a numerical approach. We therefore develop here a simpler linear approximation that is much
easier to handle and yet able to provide a physical insight.

The plan of the paper is  as follows. In Section
\ref{caseEF}, we define the tidal force field acting on the whole
LG.  The tidal forces are those developed by external groups of
galaxies. We lump together the MW and its satellites (M31 and its
satellites) for which a suitable treatment is required and look at
the remaining dwarf galaxies. On a long time scale the external tidal
forces can engender a planar distribution of these dwarf galaxies of
the LG. In Section \ref{conse_casu} we go deeper into this issue and
check whether the planar distribution is a mere coincidence or the
consequence of fundamental laws of mechanics. The answer is the latter: the
planar distribution corresponds to a minimum energy and stable
configuration of the whole system. The dwarf galaxies must lie on a
plane as a consequence of the long time scale influence of the tidal
forces exerted by massive galaxies or galaxy groups external to the
LG. Furthermore, the plane found by PC07 and the minimum energy
plane are coincident and the situation is stable. Finally, in
Section \ref{conclusion} we summarize the results and present some
general considerations.

\begin{table}
\caption{External galaxy groups gravitationally influencing the LG
\mbox{dynamics}. Each group is indicated by the dominant galaxy. The
members of each group are the same as in PC07 to whom the reader
should refer for all details (their Section 4 and Table 2). To this
list the MW and M31 are added. The radial velocities are quoted relative to the
center of the MW. No correction for the motion of the Sun toward the
Local Standard of Rest is applied because it falls below the
accuracy adopted in this study. The uncertainties on the distances
and radial velocities are omitted.}
 \centerline { \begin{tabular}{|lrrccr|}
\hline
Group & $l$ & $b$ & $d$ & $M$ & $ V_{\text{r}}$ \\
Name & $^{\circ}$ & $^{\circ}$ & Mpc & $10^{12} M_\odot$ & km~s$^{-1}$ \\
 \hline
 IC 342    & 138.2 &  10.6 & 3.3 & 12.6  &   171.0 \\
 Maffei    & 136.4 &  --0.4 & 3.5 &  6.3  &   152.0 \\
 M 81      & 142.1 &  40.9 &  3.7 &  1.6  &   130.0 \\
 Cen A     & 309.5 &  19.4 &  2.7 &  4.7  &   371.0 \\
 Sculptor  & 105.8 &  85.8 &  3.2 &  6.3  &   229.0 \\
 M 83      & 312.1 &  25.9 &  4.5 &  0.8  &   249.0 \\
 \hline
 Andromeda & 121.21 & --21.60 & 0.76 &  3.16 & --123.00 \\
 Milky Way & 0     &     0 & 0.00  &  2.20  &     0.00 \\
 \hline
\end{tabular}}
\label{Tabella01}
 \end{table}

\section{LG structure and tidal forces }\label{caseEF}
Looking at the composition of the LG, three main components can be
identified: two massive galaxies (MW and M31), their respective
groups of bounded satellites, a large number of  distant dwarfs galaxies loosely interacting or even uncoupled to
the dominant galaxies. As far as the gravitational interaction is
concerned, there are two questions we are interested in addressing:

(1) How does the external force field change with time?

(2) Have the dwarf galaxies that are not members of the MW or the M31
family been affected by the tidal interaction with the external
force field?

First, to get  a rough estimate of the influence of the
tidal forces acting on the LG, we must develop an accurate
geometrical description of the LG during its temporal
evolution. Although the tidal forces are weak, they can produce a cumulative effect of compression that could explain the planar distribution by acting over a very long time. However, if this action is
not always pointing in the same direction, the net effect can
be small, and the tidal effect can no longer be the physical
mechanism compressing the dwarf galaxy distribution. To follow the
direction of a tidal field, we apply the usual formalism of the tidal
tensor (e.g. \cite{1973grav.book.....M}, Chap. 1) to all galaxies of
the the LG and try to understand its global behavior under the action of the
external force field.

\begin{figure}
\resizebox{\hsize}{!}{\includegraphics{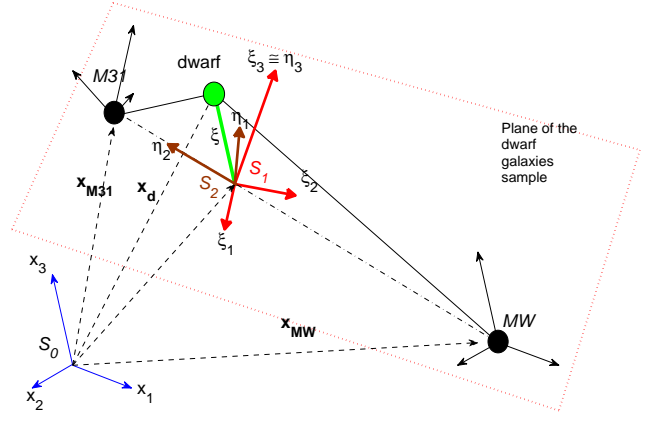}}
\caption{Sketch of the geometrical framework  we have
adopted. First we define the inertial reference frame, always named $S_0$,
with axis $\left( {S_0 ,x_1 ,x_2 ,x_3 } \right)$.
It is centered on the
CM of the external galaxy groups listed in Table
\ref{Tabella01}. We then introduce
two auxiliary reference frames: the first centered on the
barycenter of the MW and M31, which is aligned with the
principal axes of the eigen-system provided by Eqn. \eqref{E19}. This system  is
called $S_1 $ and it has axes $\left( {\xi _1 ,\xi _2 ,\xi _3 }
\right)$ with $\xi _3 $ pointing roughly in the direction orthogonal
to the plane of the dwarf galaxies (red dotted rectangle). The second is
called $S_2 $ and it has axes $\left( {\eta _1 ,\eta _2 ,\eta _3 }
\right)$ with $\eta _3 $ pointing in the same direction of $\xi _3 $ and $\eta _2 $ pointing in the direction of M31; see the
text for details. The current position vector of a generic dwarf
galaxy in the system $S_1 $ and $S_2 $ is called ${\bm{\xi }}$ and ${\bm{\eta }}$ respectively (e.g. we show ${\bm{\xi }}$ in the figure). It is
always oriented  toward  the generic dwarf galaxy. }
\label{Figure02}
\end{figure}

\textbf{Frame of reference}. The actual mutual interaction between the MW and M31 suggests
that they can be considered as a privileged system whose center of
mass (CM) can be assumed as the origin of a reference frame
(presented in Fig. \ref{Figure02} and described in more detail
below). The positions and motions of any other dwarf of
the LG that is not a member of the MW and M31 families can be given relative to this CM. This is
reminiscent of the geometry of the restricted 3 body problem (e.g.
\citet{1967torp.book.....S}) but here the CM will be followed in its time evolution. In particular the stationary action principle has been applied in PC07 to produce a possible  solution for the motion of the external groups acting gravitationally on the LG as well as MW and M31 (their table 4). From this solution we  can infer the spatial evolution in the time, $t$, of the entire LG-barycentric system ${\bm{x}}^{{\text{CM}}_{{\text{LG}}} }  = {\bm{x}}^{{\text{CM}}_{{\text{LG}}} } \left( t \right)$, which is nearly coincident with the center of mass of the MW +
M31 system.

Here, we start by considering the general expression for the
external tidal tensor due to any potential $\Phi$ expressed in the reference system $S_0$ of Figure \ref{Figure02}, to express the
external tidal tensor acting on the complex MW+M31
\[
T_{{\text{ij}}} \left( {{\bm{x}}^{{\text{CM}}_{{\text{LG}}} } } \right) = \sum\limits_{{\text{g}} \ne {\text{MW,M31}}} {\frac{{GM_{\text{g}} }}
{{\left\| {x_{\text{i}}^{{\text{CM}}_{{\text{LG}}} }  - x_{\text{i}}^{\text{g}} } \right\|^3 }} \times }
\]
\begin{equation}\label{E19}
\left( {\frac{{3\left( {x_{\text{i}}^{{\text{CM}}_{{\text{LG}}} }  - x_{\text{i}}^{\text{g}} } \right)\left( {x_{\text{j}}^{{\text{CM}}_{{\text{LG}}} }  - x_{\text{j}}^{\text{g}} } \right)}}
{{\left\| {x_{\text{i}}^{{\text{CM}}_{{\text{LG}}} }  - x_{\text{i}}^{\text{g}} } \right\|^2 }} - \delta _{{\text{ij}}} } \right)
\end{equation}
where $G$ is the gravitational constant, $x^{\text{g}}$ are the coordinates of
the external galaxy groups of Table \ref{Tabella01} of mass $M_\text{g}$,
$\delta$ is the Dirac's function, $\|.\|$ is the standard norm. We
evaluate the tidal tensor  at the barycenter of the LG, ${\bm{x}}^{{\text{CM}}_{{\text{LG}}} }  = {\bm{x}}^{{\text{CM}}_{{\text{LG}}} } \left( t \right)$ as defined above but where the time dependence has been omitted.

\begin{table*}
\caption{This table shows the temporal evolution of the eigenvalues
and eigenvectors as a function of the look-back time in Gyr (column
1). The three eigenvalues are in columns (2) to (4) and the
corresponding eigenvectors are in columns (5) through (13). The
headers are self explanatory. We have highlighted the most negative
values (column 4) and the corresponding eigenvectors (column 11
through 13) for all values of the look-back time at  which the most
negative eigenvalue keeps its  sign.}
 \centerline {\begin{tabular}{|crrrrrrrrrrrr|}
 \hline
$t_{\text{lb}}$ & $\lambda _1 $ & $\lambda _2 $ & $\lambda _3 $ & $n_{\text{x}}^{\left( {\lambda _1 } \right)} $ &  $n_\text{y}^{\left( {\lambda _1 } \right)} $ &  $n_\text{z}^{\left( {\lambda _1 } \right)} $ & $n_\text{x}^{\left( {\lambda _2 } \right)} $ &  $n_\text{y}^{\left( {\lambda _2 } \right)} $ &  $n_\text{z}^{\left( {\lambda _2 } \right)} $ & $n_\text{x}^{\left( {\lambda _3 } \right)} $ &  $n_\text{y}^{\left( {\lambda _3 } \right)} $ &  $n_\text{z}^{\left( {\lambda _3 } \right)} $ \\
\hline
13.3 &  0.0125 & -0.0049 & -0.0076       &  0.310 &  0.507 &  0.804 &  0.934 & -0.008 & -0.356 &  0.174      & -0.862       &  0.476 \\
12.3 & 0.0058 &  0.0001 &  -0.0060 & -0.266      &  0.411       &  0.872       &  0.944 &  0.292 &  0.151 &  0.192 & -0.864 &  0.465  \\
11.0 & 0.0031 &  0.0008 &  -0.0038  &  0.941      &  0.052       & -0.334      &  0.265 &  0.498 &  0.826 &  0.209 & -0.866 &  0.455 \\
9.5  &  0.0030 & -0.0002 & {\it -0.0028} &  0.971 &  0.235 & -0.043 & -0.066 &  0.436 &  0.897 & {\it 0.230} & {\it -0.868}   &  {\it 0.439} \\
7.9  &  0.0030 & -0.0006 & {\it -0.0024} &  0.954 &  0.297 &  0.026 & -0.146 &  0.390 &  0.909 & {\it 0.259} & {\it -0.871} & {\it 0.416} \\
6.2  &  0.0030 & -0.0007 & {\it -0.0023} &  0.939 &  0.341 &  0.045 & -0.171 &  0.346 &  0.922 & {\it 0.299} & {\it -0.873} & {\it 0.384} \\
4.6  &  0.0033 & -0.0008 & {\it -0.0024} &  0.919 &  0.391 &  0.050 & -0.175 &  0.292 &  0.940 & {\it 0.353} & {\it -0.873} & {\it 0.336} \\
3.0  &  0.0037 & -0.0009 & {\it -0.0028} &  0.887 &  0.458 &  0.059 & -0.171 &  0.208 &  0.963 & {\it 0.429} & {\it -0.864} & {\it 0.263} \\
1.4  &  0.0046 & -0.0010 & {\it -0.0036} & -0.829 & -0.551 & -0.097 & -0.158 &  0.065 &  0.985 & {\it 0.536} & {\it -0.832} & {\it 0.141}\\
0.0  &  0.0065 & -0.0017 & {\it -0.0047} & -0.738 & -0.657 & -0.151 & -0.064 & -0.154 &  0.986 & {\it 0.671} & {\it -0.738} & {\it-0.071} \\
\hline
\end{tabular}}
\label{Tabella04}
\end{table*}

To get  an  idea of the effects we are looking for, let us make the
following preliminary considerations. Let us assume that the most distant dwarf tat define the plane $\pi $ discovered by PC07 owe
their distribution to the initial conditions determined by the tidal
tensor of external objects. Then we expect the total tidal tensor
(TTT), $T_{\text{ij}}^{\text{tot}} \left( {\bf{x}} \right)$, defined by Eqn. (\ref{E19}) including in the sum also  the MW and M31, to
give, if reduced to its normal form, the most negative eigenvalue. Between the three eigenvectors of TTT, the one corresponding to this most negative eigenvalue will indicate the direction of the
tidal compression (see classical text-books such as
\cite{1973grav.book.....M} Chap 1, \cite{1987gady.book.....B} Chap
7, or some applications as in \cite{1989MNRAS.240..195R}). Thus if
we find this behavior also in our TTT evaluated at the position of a dwarf galaxy
far away from the MW or M31 this could hint that the compression does indeed occur.
 For example a simple case would be
that a dwarf presently belonging to the plane $\pi $ was also formed
by some mechanism  in the plane $\pi $ or close to it. In such
a case we can simply make the hypothesis that the tidal force acting on this dwarf
should be similar to the tidal force that acted on the plane
in the past, say 9 Gyr ago. We can estimate these eigenvalues and their eigenvector directions by combining the
equations of the plane $\pi$ (see  Eqn. \eqref{planeequation}), obtained by looking at the current dwarf galaxy distribution, and
Eqn. \eqref{E19} which can also be evaluated backwards in time. If we find compatible values between different points at different epochs, then we can
claim that the effect we are searching for could effectively have acted.

If the plane $\pi $ is a slab with a diameter of 4 Mpc and 200
kpc thick, we can evaluate the TTT at any point on this plane
$P\in\pi$, say 2 Mpc away from the barycenter of the LG, $T_{\text{ij}}^{\text{tot}} \left( {{\bf{x}}^{\text{P}} } \right)$. The
resulting eigenvalue of this tidal tensor, e.g. 9 Gyr ago, is $\lambda _{\text{9 Gyr}}^{\text{P}}  =
\left\{ {{\rm{0}}{\rm{.024}}{\rm{,  - 0}}{\rm{.016}}{\rm{,  -
0}}{\rm{.007}}} \right\}$.
The same
evaluation can then be repeated for the barycenter position $T_{\text{ij}}^{\text{tot}} \left( {{\bf{x}}^{\text{CM}_{\text{LG}} } } \right)$, obtaining $\lambda _{\text{9
Gyr}}^{\text{CM}_{\text{LG}}} = \left\{ {{\rm{0}}{\rm{.022}}{\rm{,  -
0}}{\rm{.014}}{\rm{,  - 0}}{\rm{.007}}} \right\}$. This result strongly suggests that back in the past the force
determining the subsequent orbital evolution of a generic dwarf had
a component squeezing the motion toward the plane. Proceeding in this way we can prove the compatibility of the eigenvalues of the TTT for every position on the plane $\pi$, i.e. $T_{\text{ij}}^{\text{tot}} \left( {{\bf{x}}^{\text{CM}_{\text{LG}} } } \right) \cong T_{\text{ij}}^{\text{tot}} \left( {{\bf{x}}^{\text{P}} } \right)\forall P \in \pi $. This clearly allow us
to explore the possibility that the plane $\pi$ of PC07 is the
consequence of the tidal forces acting on the LG during a large
fraction of the Hubble time, i.e. we want to extend this estimation not only to the present time $t=t_0$ but also to the time\footnote{Of course a more correct computation could have been performed by knowing distribution of the dwarf galaxies in the past, but unfortunately we cannot track back the past orbits of the dwarf galaxies belonging to the plane $\pi$ today; therefore Eqn. \eqref{planeequation} for the plane $\pi$ cannot be directly determined in its time evolution.} $t<t_0$.

\begin{figure}
\resizebox{\hsize}{!}{\includegraphics{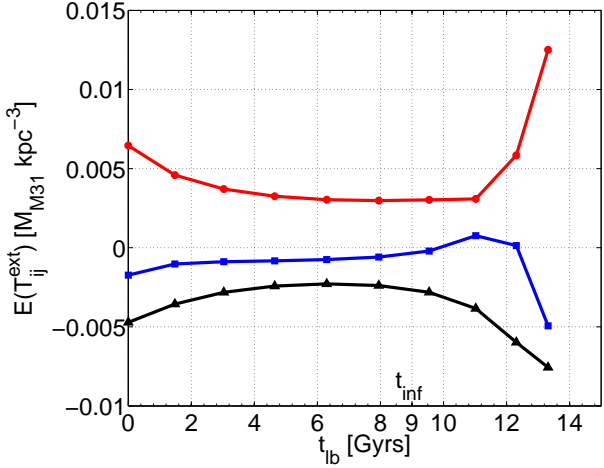}}
\caption{Evolution of the eigenvalues $E(T_{\text{ij}}^{\text{ext}})$ of Table \ref{Tabella04} as a
function of the look-back time in Gyr (column 1). The red  line with
dots is for the first positive eigenvalue $\lambda_1$, the blue line
with squares is  the second eigenvalue $\lambda_2$ negative for $t_{\text{lb}}<9.5Gys$, and
the black with triangles is for the third (most negative) eigenvalue $\lambda_3$. }
\label{Quadrupolo_GL_autovalori_evoluzione}
\end{figure}

To proceed further, we need to search   the  eigenvectors
associated with the tidal tensor with the most negative eigenvalues.
In the limits of our approximation, they should keep a
direction with respect to an inertial reference frame not too far from the normal to the geometrically plane ${\bm{\hat n}}\left( \pi  \right)$ for most of
the Hubble time, which can nowadays be inferred by simple inspection of the dwarf galaxies' distribution in the LG. To prove this, we solve  the eigen-system for the
tidal tensor of Eqn. \eqref{E19} as a function of time. The solutions are presented in
Table \ref{Tabella04}. The results we are interested in are limited in time to a range where monotonicity of the trend of the eigenvalues can be exploited in order to reveal an integrated cumulative effect of compression or expansion. We find that our range of interest has to span the last 9 Gyr, imposing a lower limit to our analysis of $t=t_{\text{inf}}\cong9 $ Gyr. Before this $t_{\text{inf}}$ the configuration of the eigenvalues is slightly different. From Table \ref{Tabella04} we see that the time evolution of the eigenvalues shows a phase with one compressive direction compared to two positive expansion directions, and before that, as well as after $t=t_{\text{inf}}$, we see two negative directions compared with one positive. For simplicity we will not be treating analytically in the following sections these switchings between the configurations. We are only interested in the last most dominant time evolution of the monotonic behavior of the eigenvalues.
As we will exclude the first three rows of Table \ref{Tabella04} (the primordial evolution prior to $t_{\text{inf}}$) from our analysis, from now on we will uniquely identify (unless otherwise specified) with $\lambda_1$, $\lambda_2$, $\lambda_3$ the positive eigenvalue, the second negative eigenvalue and the most negative eigenvalue respectively for the eigenvectors of the external tidal tensor defined in equation \eqref{E19} followed in their time $t$ evolution exclusively for $t \in \left] {t_{\text{inf} }, t_0} \right]$. The most negative eigenvalue (column 4) and its evolution during the
past 9 Gyr  and  projections of the associated eigenvector onto the
axes of the inertial system are highlighted in \textit{italics}. The
time variations of the three eigenvalues are shown in Fig.
\ref{Quadrupolo_GL_autovalori_evoluzione}.

\begin{figure}
\resizebox{\hsize}{!}{\includegraphics{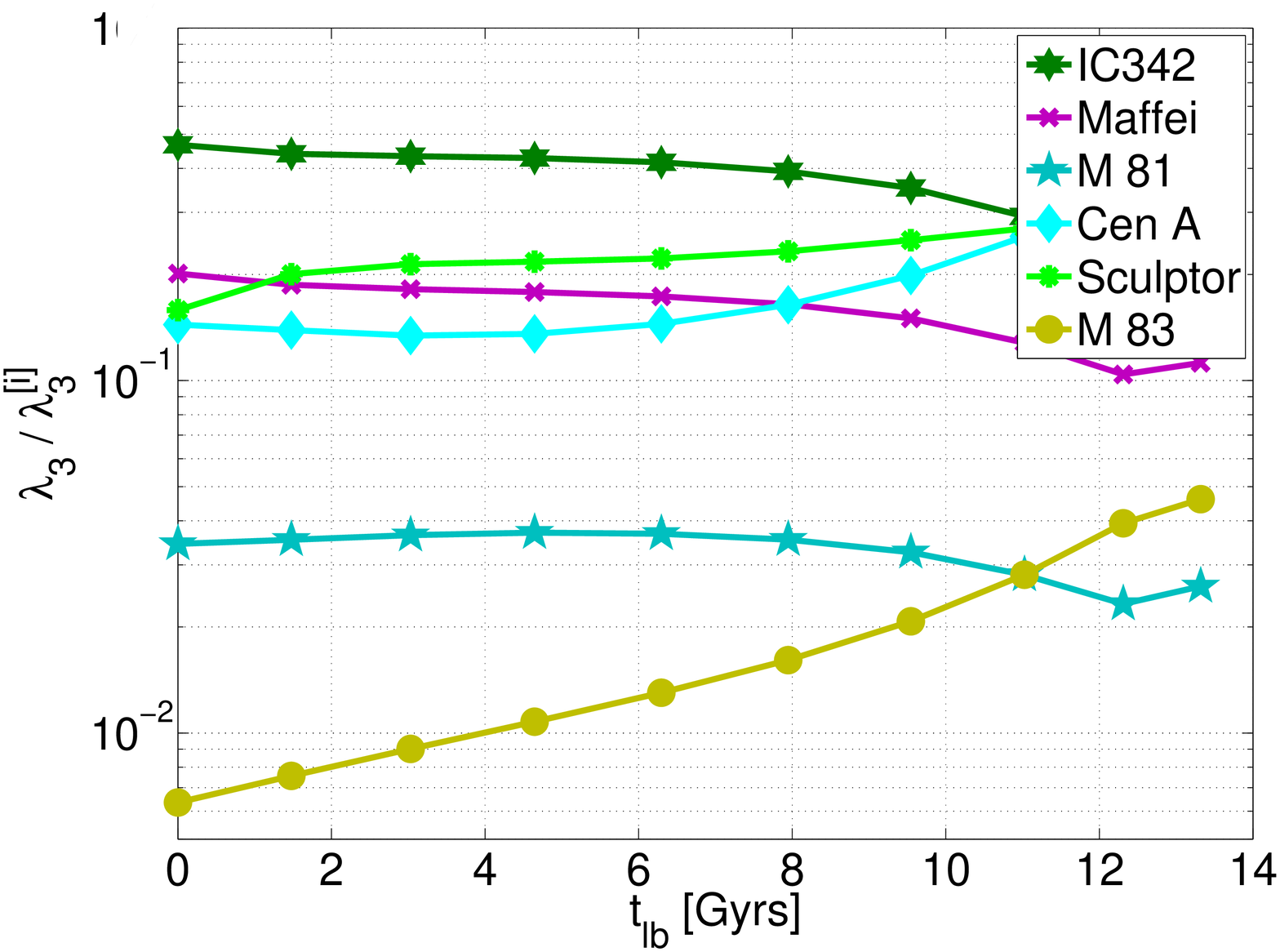}}
\caption{Here we tracked back the influence of each individual component in the sum of Eqn. \eqref{E19} for the most negative eigenvalue, $\lambda_{3}^{[\text{i}]}$. The curves  are normalized to the black line, $\lambda_{3}$, of Fig. \ref{Quadrupolo_GL_autovalori_evoluzione} (see text for details). The major contribution overall is dominated by the IC342 group, followed by Cen A, Maffei and Sculptor groups. The contribution of M83, even if nowadays marginal, could have played a role in the past in the imprint of the initial condition of the LG dwarf galaxies.}
\label{Quadrupolo_GL_autovalori_negativo}
\end{figure}

It is soon evident from the data displayed in  Fig.
\ref{Quadrupolo_GL_autovalori_evoluzione} that one of the negative
eigenvalues  dominates. Therefore, we are  in a situation in which
the external field acting on the LG may engender a planar
distribution of the dwarf galaxies. This is an interesting result
because

\begin{enumerate}
    \item It proves that,  using the tidal tensor,
    we can  analytically obtain the same
results of PC07 for the behavior of the external field.
 The external force field turns out to be
compatible  with a flat spatial distribution of the dwarf galaxies
that remains constant for a large fraction of the Hubble
time. Moreover, we can argue  that the external force field
started  to flatten  the spatial distribution of the dwarf galaxies
already prior to $t_{\text{inf}}$. If we investigate in more detail the relevance of the different groups on this flattening effect, we can plot in Fig. \ref{Quadrupolo_GL_autovalori_negativo} the normalized trend of the most negative eigenvalue of Eqn.\eqref{E19} split into its components. Here we see the most negative eigenvalues of the sum in  Eqn.\eqref{E19} normalized to the overall sum (hence the black line of Fig. \ref{Quadrupolo_GL_autovalori_evoluzione} is here the unitary constant upper bound of the figure). As we can infer from the figure, the influence of IC342 group has always been the most significant, followed by the effects of Maffei, Sculptor and Cen A groups. This is expected from their masses and positions listed in Table \ref{Tabella01}. The time evolution shown in this Figure confirms their relative importance in the compressing effect on the LG for its temporal evolution. Slightly less important is the contribution of the M83 group that was nevertheless as important as that of the M81 group 11 Gyr ago.

    \item  We must clarify once and for all that the present result does \textit{not}
prove that the spatial distribution of dwarf galaxies in  the LG has
to be  flat, but   only that the external force field is compatible
with such a flat distribution.

\item The fact that the external force field acting  only on the two
main HGs (MW and M31) of the LG is compatible with a flat spatial
distribution of the distant dwarfs does not tell us anything about the
distribution of the nearby dwarf satellites around their HG (both MW and
M31).

\item  Furthermore, the planar distribution shown by the geometrical
analysis made by PC07 cannot \textit{a priori } be related to the planar
distribution suggested by the compression effect described by the tidal
tensor affecting the LG. The subject of the following analysis is to explain the coincidence claimed by PC07 between
the geometrical and dynamical planes.
\end{enumerate}

Now we seek to prove that the following two issues are
tightly related: (i) There exists a plane in the spatial
distribution of the LG dwarf galaxies  that is expected from the external
tidal force acting on the LG. (ii) In the context of the linear
approximation that we have adopted (see below), the tidal force field
compatible with a flat distribution has an orientation whose normal
vector is compatible with the normal vector of the planar
distribution  $\pi$ found by PC07.

\begin{figure}\label{QinCtuttituttiiTGL2}
\resizebox{\hsize}{!}{\includegraphics{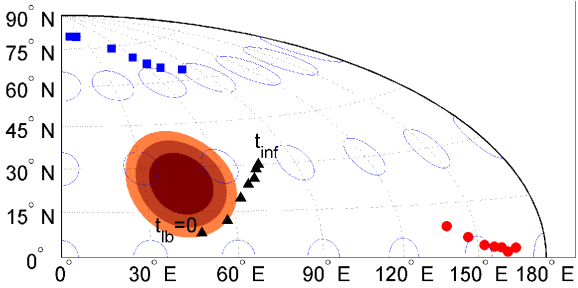}}
\caption{The Hammer projection of the sky position of the normal
to the  plane found by PC07 (from purely geometrical arguments). The color code from light red to dark brown corresponds to
the 3$\sigma$, 2$\sigma$, and 1$\sigma$ uncertainties as estimated by
PC07 from the principal component analysis and projected onto the
sky. The shape of the three shaded areas becomes more oblate at
increasing angular direction \textit{b}. The classical Tissot's circles, perfect circles of angular radius of $7$ deg have been plotted to help visualize the angular distances. We have used the same symbols and time intervals as in Fig.
\ref{Quadrupolo_GL_autovalori_evoluzione} and Table \ref{Tabella04}. The black triangles
nearly overlap the shaded area, showing that  the direction of the
eigenvector with the  most negative eigenvalue lies close to the direction of the normal to the PC07 plane, which is of course constant in the figure because it is derived from the observational data available today. For comparison
we  also show the evolution of the other two orthogonal eigenvectors
indicated by the same symbols as in Figure \ref{Quadrupolo_GL_autovalori_evoluzione}.}
\end{figure}

The equation for the plane $\pi$   can be rewritten here in
the reference frame  $S_0$  as

\begin{equation}\label{planeequation}
    0.64x_1 - 0.61x_2 - 0.45x_3  =0
\end{equation}

\noindent
with a direction $\left( {l,b} \right) = \left( { 45°,
27°} \right)$ for ${\bm {\hat n}}\left( \pi \right)$, the normal
vector to the plane. Furthermore, the directions of the three basis
vectors ${\bm {\hat e}}_{\text{S}_0 } $ in which the eigenvalues of Table
\ref{Tabella04} are expressed have been  chosen as collinear with
those of the reference frame adopted by PC07 (the approximation was
made such that the Sun is placed at the center of MW, the error of $
\approx 8.5$ kpc is negligible in the present context). Therefore, we
can assume that \textit{only} at the present time $t_0$ does the
reference system in which the above equation of the plane is
written have orthonormal vectors parallel to the basis vector of
reference system adopted at $t=t_0$. It follows that the
Hammer projection from the center of mass of the LG shows the eigenvector directions
 as a function of time and the normal vector to the geometrical plane
of PC07 are as in Fig. \ref{QinCtuttituttiiTGL2}.

Interestingly, it is the eigenvector relative to the most negative eigenvalue that lies closer to the direction of the normal of the plane of the dwarf galaxies.

Considering that every point in this map has an error radius of about
$ \pm 9^\circ $, inherited from the Minimum Action analysis, and
considering the uncertainty in the angular definition of the
direction of ${\bm{\hat n}}\left( \pi \right)$, \textit{the
normal to the geometrical plane and the direction of the vector
associated with the most negative eigenvalue are compatible at the present time at
the 3$\sigma$-level of confidence}. Moreover, the eigenvector of the most negative
eigenvalue  $\lambda_3 $ of the tidal tensor moves, maintaining a direction not so far form the direction that we can nowadays deduce from the observation for ${\bm{\hat n}}\left( \pi  \right)$.
 This result, which was already present  in PC07, is recovered
here in a semi-analytical treatment of the whole problem.
\textit{The key question to be answered now is whether the
coincidence is "causal  or casual" and what the meaning
of all this is}.

\section{Causal or casual?}\label{conse_casu}
 The tidal tensor  can be
derived from the Taylor expansion of the force field (see next
section). This implies that the object we want to investigate (a
dwarf galaxy) is under the effect of a smoothly varying potential in
the course of evolution.

The
large scale description of the gravitational interaction adopted here and by
PC07 does not work at the distance scales of the closer HG-dwarf
satellite interactions where satellite dwarf
galaxies suffer kick-off, are continually absorbed into the halo of
the HG (MW and M31 in our case) and the collisionless
description is not correct. A dwarf galaxy of these
closer samples, undergoing the much more intense direct
interaction with the HG, does not significantly respond to  the
weaker external field acting on it.
On the other hand, the collisionless description is
suited to deal with the effects of distant galaxy groups (see the list in
Table \ref{Tabella01}) and hence the influence of the
external field on the other  dwarf galaxies  far away  from a HG (see e.g. \citet{1989MNRAS.240..195R, 1990ApJ...362....1P, 1993MNRAS.264..865D, 1994ApJ...429...43P}).

\textbf{The frame of reference}. In order to formalize what is in the previous section we
present the following two hypotheses, one on the directions of the eigenvectors and the other on their values:

\begin{enumerate}
    \item We will not consider the evolutionary stages of the
Universe older than $t_{\text{lb}} > 9$ Gyr (look-back time) to work with a monotonic behaviour of a single negative eigenvalue.
    \item  We  assume that the eigenvector corresponding to the most negative eigenvalue always points
in the same direction with respect to the inertial reference frame
in the non-comoving coordinate system.
\end{enumerate}

First
we define  $S_0 $ as the inertial reference frame centered on the
CM of the external galaxy groups listed in Table
\ref{Tabella01}. We introduce also  a second reference frame $S_1
$, non-inertial and comoving with the CM of the LG,
approximately coincident with the CM of the pair  M31 and
MW. Moreover, we assume that this reference frame has its principal
axis always collinear  with the principal axis of the external tidal
tensor, i.e. this frame is an eigen-system of the external tidal
tensor. ${\bm{x}}_{\text{M31}} $ is the
coordinate vector of M31 in $S_0 $, ${\bm{x}}_{\text{MW}} $ is the
coordinate vector of MW in $S_0 $, ${\bm{x}} $ is the vector of
a generic dwarf galaxy in $S_0 $ and ${\bm{R}}$ is the position of
the M31-MW barycenter, i.e. the origin of $S_1 $ in $S_0 $. The
generic position vector in the frame $S_1 $ is ${\bm{\xi }}$. It
will be used to indicate the generic position of a dwarf galaxy.
Both reference frames are orthogonal (see Fig. \ref{Figure02}).

\textbf{Equations of motion}.  We only point out here  that the
generic position vectors in the two reference frames are now written
as ${\bm{x}} = {\bm{x}}_{\text{LG}}^{\text{CM}}  + {\bm{O\xi }} = {\bm{R}} + {\bm{O\xi }} $, where $\bm{O} \in SO(3)$ is the generic rotation matrix with
  elements $O_{\text{ij}}$. Similarly, for the time derivative of
${\bm{x}}$ we can write ${\bm{\dot x}} = {\bm{\dot x}}_{\text{LG}}^{\text{CM}} +
\bm{O}\left( {{\bm{\dot \xi }} + {\bm{\Omega }} \times {\bm{\xi }}}
\right) $
 or, in more compact notation
 $${\bm{ \dot x}} = {\bm{V}} + \bm{O}\left( {{\bm{\dot \xi }} + {\bm{\Omega }}
  \times {\bm{\xi }}} \right),$$
where $\bm{\Omega}$ is the angular velocity and we put
${\bf{V}} = {\bf{\dot x}}_{\text{LG}}^{\text{CM}} $.

The time derivative of the velocity equation yields the
  equation of motion

    \[
 {\bm{\ddot x}} = {\bm{A}} + \bm{O}\left( {{\bm{\ddot \xi }}  + 2{\bm{\Omega }} \times {\bm{\dot \xi }} + {\bm{\dot \Omega }} \times {\bm{\xi }} + {\bm{\Omega }} \times \left( {{\bm{\Omega }} \times {\bm{\xi }}} \right)} \right), \\
\]
where, in the second equation the vector ${\bm{A}}$  has been used to
simplify the notation for the acceleration ${\bm{\ddot x}}_{\text{LG}}^{\text{CM}}
$ of the barycenter.
Moreover, to better  underline the physical  meaning of the different terms
we recollect the previous equation as
\begin{equation}\label{E21}
    {\bm{\ddot \xi }} = \bm{O}^T \left( {{\bm{\ddot x}} - {\bm{A}}} \right) - 2{\bm{\Omega }} \times {\bm{\dot \xi }} - {\bm{\dot \Omega }} \times {\bm{\xi }} - {\bm{\Omega }} \times \left( {{\bm{\Omega }} \times {\bm{\xi }}} \right).
\end{equation}
From this last equation it is quickly evident that the acceleration $\bm{\ddot \xi }$, suffered
by a dwarf galaxy in the non-inertial reference frame, is the sum of
different terms due to the motion of the frames $S_1 $ and
 $S_0 $  and the fictional forces that appear thanks to the non-inertial
 nature of $S_1 $:  $
 - {\bm{\Omega }} \times \left( {{\bm{\Omega }} \times {\bm{\xi }}} \right)$
 is the centrifugal effect,
 $
- 2{\bm{\Omega }} \times {\bm{\dot \xi }}$ the Coriolis effect
proportional to the velocity of the dwarf,  and $-{\bm{\dot \Omega }}
\times {\bm{\xi }}$ is the effect caused by the non-constant rotation rate of  $S_1 $.
 As usual we assume
that the rotation matrix linking the different orthonormal
reference frames is  $\bm{O}: \bm{OO}^T  = \bm{I}$ where  the $T$  stands for
transpose and $\bm{I}$ is the identity matrix.

From  Eqn. \eqref{E21} it is easily evident how to evaluate the term
which gives the acceleration of  a generic  dwarf, ${\bm{\ddot x}} -
{\bm{A}}$. Now the acceleration of a dwarf galaxy in $S_0 $ is due
to three contributions:  the gradient in the external potential
$\nabla \Phi _{\text{ext}} $, the gradient in the MW potential $\nabla \Phi
_{\text{MW}} $, and the gradient in the M31 potential, $\nabla \Phi _{\text{M31}}
$, i.e.

\begin{equation}\label{E22}
    {\bm{\ddot x}} =  - \nabla _{\bm{x}} \left( {\Phi _{\text{ext}}  + \Phi _{\text{MW}}  + \Phi _{\text{M31}} } \right).
\end{equation}
Taylor expanding $\Phi_{\text{ext}}$ at the second order around the comoving barycenter, and with the usual definition of the Tidal Tensor given
in equation \eqref{E19}, we can easily write

\begin{equation}
\begin{gathered}
  \Phi _{\text{ext}} \left( {\bm{x}} \right) \simeq \Phi _{\text{ext}} \left( {\bm{R}} \right) + \left\langle {\nabla _{\bm{R}} \Phi _{\text{ext}} \left( {\bm{R}} \right),{\bm{x}} - {\bm{R}}} \right\rangle  -  \\
  \frac{1}
{2}\left\langle {{\bm{T}}_{\text{ext}} \left( {\bm{R}} \right)\left( {{\bm{x}} - {\bm{R}}} \right),{\bm{x}} - {\bm{R}}} \right\rangle  \\
\end{gathered}
\end{equation}
together with  two other similar relations for the potentials of MW
and M31 that are not given here for the sake of brevity, with
$\langle.,.\rangle$ expressing the standard inner product. Inserting
these three relations in Eqn. \eqref{E22}, after some simplifications
we obtain  the acceleration ${\bm{\ddot x}} - {\bm{A}}$ in linear
approximation:
\[
{\bm{\ddot x}} - {\bm{A}} \simeq \left\langle {{\bm{T}}_{\text{ext}} \left(
{\bm{R}} \right) + {\bm{T}}_{\text{MW}} \left( {\bm{R}} \right) +
{\bm{T}}_{\text{M31}} \left( {\bm{R}} \right),{\bm{x}} - {\bm{R}}}
\right\rangle.
\]

\noindent By  introducing  the TTT as defined in the preceding section with
$${\bm{T}}\left( \bm{R}
\right) = {\bm{T}}_{\text{ext}} \left( {\bm{R}} \right) + {\bm{T}}_{\text{MW}}
\left( {\bm{R}} \right) + {\bm{T}}_{\text{M31}} \left( {\bm{R}} \right)$$
and using ${\bm{x}} - {\bm{R}} = \bm{O}{\bm{\xi }}$, we can  simplify Eqn.
\eqref{E21} that becomes

\begin{equation}\label{E24}
    {\bm{\ddot \xi }} = {\bm{O}}^T {\bm{TO\xi }} - 2{\bm{\Omega }} \times {\bm{\dot \xi }} - {\bm{\dot \Omega }} \times {\bm{\xi }} - {\bm{\Omega }} \times \left( {{\bm{\Omega }} \times {\bm{\xi }}} \right),
\end{equation}
where we have dropped the explicit dependence on the position of
 the tidal tensor  ${\bm{T}}$. This tensor is always  evaluated at
the barycenter ${\bm{T = T}}\left( {\bm{R}} \right)$ if not specified otherwise.

\subsection{ Energy states of equilibrium}
Our aim now is to understand whether the equation of motion
\eqref{E24} can lead to stable equilibrium configurations. There are
several techniques for attacking  this problem based on the
integration over a time interval of the force or the impulse, see
e.g. \cite[ Chap 7]{1987gady.book.....B}, the elegant Lagrangian
treatment of \cite{1999ApJ...514..109G}, or the sophisticated
analysis  in the action space of \citet{1994AJ....108.1398W,
1994AJ....108.1403W}.

To proceed further we look for the energy equilibrium configurations
and their stability of a system  governed by the equations of motion
\eqref{E24}. Here, we take advantage of the fact  that we can follow
the evolution of the angular velocity of the non-inertial reference
frame by looking at  the motion on the sky of the eigenvector of the
tidal tensor as already made in previous sections for the quadrupole,
see Fig. \ref{Quadrupolo_GL_autovalori_evoluzione}.

It can be demonstrated  that the Lagrangian  leading to Eqn.
\eqref{E24}, up to a total derivative, can be written as
\[
L = \frac{m}{2}\left\| {{\bm{\dot \xi }}} \right\|^2  +
m\left\langle {{\bm{\dot \xi }},{\bm{\Omega }} \times {\bm{\xi }}}
\right\rangle  + \frac{m}{2}\left\| {{\bm{\Omega }} \times {\bm{\xi
}}} \right\|^2  - m\left\langle {{\bm{A}},{\bm{O\xi }}}
\right\rangle  - W
\]
 \cite[see for instance][]{1969mech.book.....L}. Therefore, taking  the linear
 momentum ${\bm{p}} = \frac{{\partial L}}{{\partial {\bm{\dot \xi }}}} =
m{\bm{\dot \xi }} + m\left( {{\bm{\Omega }} \times {\bm{\xi }}}
\right)$
 we can evaluate the energy $ E = \left\langle
{{\bm{p}},{\bm{\dot \xi }}} \right\rangle  - L$
 as
    \[
E = \frac{m}{2}\left\| {{\bm{\dot \xi }}} \right\|^2  -
\frac{m}{2}\left\| {{\bm{\Omega }} \times {\bm{\xi }}} \right\|^2  +
m\left\langle {{\bm{A}},{\bm{O\xi }}} \right\rangle  + W.
\]
Taking  the derivative with respect to the positions, we
 obtain
\begin{equation}\label{E25}
    \frac{{\partial E}}{{\partial {\bm{\xi }}}} =
    m{\bm{\Omega }} \times \left( {{\bm{\Omega }} \times {\bm{\xi }}} \right) - m{\bm{O}}^T {\bm{TO\xi}}.
\end{equation}
The equilibrium energy is then given by the solution of the equation
\begin{equation}\label{E26}
    {\bm{\Omega }} \times \left( {{\bm{\Omega}} \times {\bm{\xi }}} \right) - {\bm{O}}^T {\bm{TO\xi }} = 0.
\end{equation}
In the reference frame $S_1 $, by definition the external tidal
tensor is always diagonal,  i.e.  ${\bm{O}}^T {\bm{T}}_{\text{ext}} \left(
\bm{R} \right){\bm{O}}$ is greatly simplified but,
in contrast, the first term of Eqn. (\ref{E26}) has a complicated
structure and the same occurs for the other two terms composing ${\bm{T}}$.
This  way of proceeding does not allow for significant  simplifications but a better insight can be gained by moving to a new reference frame, $S_2$, with the same origin as the previous reference system $S_1$, in which the
matrices representing the tidal tensors ${\bm{T}}_{\text{M31}} $ and
${\bm{T}}_{\text{MW}} $ are both in diagonal form, but where ${\bm{T}}_{\text{ext}} $ is
not. From the analysis of the orbit evolution expected in a
statistical interpretation of the Minimum  Action, PC07 showed that
MW and M31 are roughly coplanar to the external force field,  at
least during the last  9 Gyr. Therefore, in our simple model we can
assume that the angular velocity vector ${\bm{\Omega }}_{\xi _3 }$
with which we describe the rotation of the system of reference
$S_1$,
 which we remember here again for sake of clarity has its third axis
  $\xi_3$ parallel to the normal ${\bf{\hat n}}\left( \pi  \right)$,
   has to be parallel to the angular velocity vector in this new
   system of reference $\left( {O,{\bm{\hat e}}_{\eta _1 } ,{\bm{\hat e}}_{\eta _2 } ,{\bm{\hat e}}_{\eta _3 } } \right)$, namely
   ${\bm{\Omega }}_{\eta _3 } \left( {S_2 }
\right)$, that we assume rotating around its third axis $\eta_3$,
 i.e. ${\bm{\Omega }}_{\eta _3 } \left( {S_2 }
\right)||{\bm{\Omega }}_{\xi _3 } \left( {S_1 } \right)$ over
the last 9 Gyr. This means that we do not allow the orbital plane of
M31 and MW to tilt with respect to the plane
$\tau _{\lambda _3 } \cong \pi$ orthogonal to the eigenvector corresponding to the most
negative eigenvalue $\lambda_3(t)$ in the course of time evolution.
Even though this assumption is automatically
fulfilled in the analysis below, before looking at the
evolution  of  the MW and M31 in the sky centered on the LG barycenter,
we have checked that the angular distance between the position
vectors of MW or M31 and ${\bm{\hat n}}\left( {\tau _{\lambda _3
} } \right)$  is $90^\circ  \pm 3^\circ $, thus indirectly
showing that  the propagation of numerical errors in the calculation
of the temporal evolution of the eigenvectors is small.

On the basis of these considerations, we can suppose that $\exists\, \bm{N}
= \bm{N}\left( t \right):{\bm{\xi }} = \bm{N}{\bm{\eta }}$, i.e. there
 exists a linear operator represented by a rotational matrix
${\bm{N}}\in SO(3)$ such that the generic position vector of a
dwarf galaxy in $S_1$, $\bm{\xi}$, can be written as a function of
the generic position vector in $S_2$, $\bm{\eta}$, and the tidal
tensor of the external potential
can be written as:
\[
\begin{gathered}
  {\bm{O}}^T {\bm{T}}_{\text{ext}} {\bm{ON\eta }} = \left( {\begin{array}{*{20}c}
   {\lambda _1 } & 0 & 0  \\
   0 & {\lambda _2 } & 0  \\
   0 & 0 & {\lambda _3 }  \\

 \end{array} } \right)\left( {\begin{array}{*{20}c}
   {N_{11} } & {N_{12} } & 0  \\
   {N_{21} } & {N_{22} } & 0  \\
   0 & 0 & 1  \\

 \end{array} } \right)\left( {\begin{array}{*{20}c}
   {\eta _1 }  \\
   {\eta _2 }  \\
   {\eta _3 }  \\

 \end{array} } \right) \\
   = \left( {\begin{array}{*{20}c}
   {N_{11} \eta _1 \lambda _1  + N_{12} \eta _2 \lambda _1 }  \\
   {N_{21} \eta _1 \lambda _2  + N_{22} \eta _2 \lambda _2 }  \\
   {\eta _3 \lambda _3 }  \\

 \end{array} } \right) \\
\end{gathered}
\]
where we have adopted a rotation matrix spinning about
$\Omega_{\eta_3}$. As we can see, no explicit dependence on the
rotational coefficients of the matrix $\bm{O}$ is necessary in this
reference system $S_2$ that we have chosen but the reader should keep in
mind that ${\bm{O}}\left( t \right) \ne {\bm{N}}\left(
t \right)\forall t$ and only in this new reference system $S_2$
have we been able to simplify the matrix as above. The general form of
the rotational matrix could  be casted as  a combination of
trigonometric functions, but this would be superfluous here. The
only important thing to note is that even if the assumption of
collinearity between the axis of rotation of $S_1 $ and $S_2 $ has
been justified, nothing can said about the moduli of their angular
velocities. We \textit{cannot} assume that the rotation of the pair
MW and M31 or, equivalently, the reference frame tightened to this
rotation spins
 with the same rotational velocity of the reference frame attached to
the external potential! This could lead to wrong or paradoxical
results that need to be avoided. In other words we cannot  assume
$\left\| {{\bm{\Omega }}_{\eta_3} \left( {S_2 } \right)} \right\| =
\left\| {{\bm{\Omega }}_{\xi_3} \left( {S_1 } \right)} \right\|$,
which is wrong, but simply adopt the condition on the directions
\[\frac{{{\bm{\Omega }}_{\eta_3} \left( {S_2 } \right)}}
{{\left\| {{\bm{\Omega }}_{\eta_3} \left( {S_2 } \right)} \right\|}} =
\frac{{{\bm{\Omega }}_{\xi_3} \left( {S_1 } \right)}} {{\left\|
{{\bm{\Omega }}_{\xi_3} \left( {S_1 } \right)} \right\|}}.
\]
For example we can explicitly write the matrix ${\bm{N}}$
 as
\[
\left( {\begin{array}{*{20}c}
   {N_{11} } & {N_{12} } & 0  \\
   {N_{21} } & {N_{22} } & 0  \\
   0 & 0 & 1  \\
\end{array}} \right) = \left( {\begin{array}{*{20}c}
   {\cos \gamma } & { - \sin \gamma } & 0  \\
   {\sin \gamma } & {\cos \gamma } & 0  \\
   0 & 0 & 1  \\
\end{array}} \right)
\]
where $\gamma  = \gamma \left( t \right)$ is the angle between
the two systems.  It will change as  a function of  time due to
the time dependence of the angular velocity of the two frames
\[
\gamma  = \gamma \left( {\Omega _{\xi _3 }^{S_1 } \left( t
\right),\Omega _{\eta _3 }^{S_2 } \left( t \right)} \right).
\]
This also means that the angular velocity is not related to the
angular momentum in a simply way.  It will indeed be the result of the
combined action  of the centrifugal force due to the rotation of the
frame tightened to the external force field, i.e. $\bm{T}_{\text{ext}} $, and
the centrifugal force due to the rotation of the frame tightened to
the motion of M31 and MW via $\bm{T}_{\text{MW}} $ and $\bm{T}_{\text{M31}} $. We
have a double centrifugal effect: one  caused by the external field
and the other by the MW and M31 that acts with different characteristic
angular velocity (typically  time-dependent angular velocities). The
intensity of the force due to the centrifugal component can be
written as
$$\begin{array}{l}
 \left\| {\bm{N}{\bm{\Omega }}\left( {S_1 } \right)} \right\|^2 \left\| {\bm{N}{\bm{\eta }}} \right\| + \left\| {{\bm{\Omega }}\left( {S_2 } \right)} \right\|^2 \left\| {\bm{\eta }} \right\| =  \\
  = \left\| {{\bm{\Omega }}\left( {S_1 } \right)} \right\|^2 \left\| {\bm{\eta }} \right\| + \left\| {{\bm{\Omega }}\left( {S_2 } \right)} \right\|^2 \left\| {\bm{\eta }} \right\| =  \\
  = \left( {\left\| {{\bm{\Omega }}\left( {S_1 } \right)} \right\|^2  + \left\| {{\bm{\Omega }}\left( {S_1 } \right)} \right\|^2 } \right)\left\| {\bm{\eta }} \right\|, \\
 \end{array}$$
where $\left\| {\bm{\eta }} \right\|$ is the distance of the
dwarf galaxy from the moving barycenter of $S_2 $. Moreover in this
frame of reference $S_2 $  we have that $\bm{O}^T \bm{T}_{\text{MW}} \bm{ON}{\bm{\eta
}}$ and $\bm{O}^T \bm{T}_{\text{M31}} \bm{ON}{\bm{\eta }}$ are diagonal. Therefore, we
immediately have (see e.g. \citet{1973grav.book.....M}, Chap 1)
$$
\bm{O}^T \bm{T}_{\text{MW}} \bm{ON}{\bm{\eta }} = \frac{{GM_{\text{MW}} }}{{\left\| {{\bm{\eta
}}_{\text{MW}} } \right\|^3 }}\left( {\begin{array}{*{20}c}
   { - 1} & 0 & 0  \\
   0 & 2 & 0  \\
   0 & 0 & { - 1}  \\
\end{array}} \right)\left( {\begin{array}{*{20}c}
   {\eta _1 }  \\
   {\eta _2 }  \\
   {\eta _3 }  \\
\end{array}} \right)
$$
where  for simplicity we have assumed that ${\bm{\hat e}}_{\eta _2 }$ points
from the barycenter of the system  toward MW, ${\bm{\hat e}}_{\eta _3 }$
 is parallel to $\Omega _{\xi _2 } \left( {S_1 } \right)$,
 and with the third axis oriented in such a way to form a left-handed reference frame.
 In the same way we may write the analogous for
  $\bm{O}^T \bm{T}_{\text{M31}} \bm{ON}{\bm{\eta }}$.
Finally, the tidal term  in the condition for the energy
equilibrium,  Eqn. \eqref{E26}, is given by
\[
\begin{array}{l}\label{E29}
 {\bm{O}}^T {\bm{TO }} = \left( {\begin{array}{*{20}c}
   {\alpha _{11} } & {N_{12} \lambda _1 } & 0  \\
   {N_{21} \lambda _2 } & {\alpha _{22} } & 0  \\
   0 & 0 & {\alpha _{33} }  \\
\end{array}} \right) \\
\rm with\\
 \alpha _{11}  = N_{11} \lambda _1  - \frac{{GM_{\text{MW}} }}{{\left\| {{\bm{\eta }}_{\text{MW}} } \right\|^3 }} - \frac{{GM_{\text{M31}} }}{{\left\| {{\bm{\eta }}_{\text{M31}} } \right\|^3 }} \\
 \alpha _{22}  = N_{22} \lambda _2  + \frac{{2GM_{\text{MW}} }}{{\left\| {{\bm{\eta }}_{\text{MW}} } \right\|^3 }} + \frac{{2GM_{\text{M31}} }}{{\left\| {{\bm{\eta }}_{\text{M31}} } \right\|^3 }} \\
 \alpha _{33}  = \lambda _3  - \frac{{GM_{\text{MW}} }}{{\left\| {{\bm{\eta }}_{\text{MW}} } \right\|^3 }} - \frac{{GM_{\text{M31}} }}{{\left\| {{\bm{\eta }}_{\text{M31}} } \right\|^3 }} \\
 \end{array}
\]
which provides a  simple description  of the tidal effects. The
above relation can be further simplified by recalling that, from the
definition of barycenter in the reference frame in use,  we can write
\begin{equation}\label{Equetas}
	\eta _2^{\text{MW}}  =  - \frac{{\eta _2^{\text{M31}} M_{\text{M31}} }}{{M_{\text{MW}} }}.
\end{equation}
Inserting this expression into Eqn. \eqref{E26}, written in the $S_2$ frame, and using the relation
$\left\| {{\bm{\eta }}_{\text{MW}} } \right\| = \sqrt {0 + \left( {\eta
_2^{\text{MW}} } \right)^2  + 0}  = \left| {\eta _2^{\text{MW}} } \right| $ i.e.
$\left\| {{\bm{\eta }}_{\text{MW}} } \right\|^3  = \left| {\eta _2^{\text{MW}}
} \right|^3 $,  after tedious algebraic simplifications, we get for
the three components of the tidal energy
\[
{\bm{O}}^T {\bm{T{\rm O}\eta }} =
\]
\[
=\left(
{\begin{array}{*{20}c}
   {N_{12} \lambda _1 \eta _2  + \eta _1 \left( {N_{11} \lambda _1  - \frac{{G\left( {M_{\text{MW}}^4  + M_{\text{M31}}^4 } \right)}}{{M_{\text{M31}}^3 \left| {\eta _2^{\text{M31}} } \right|^3 }}} \right)}  \\
   {N_{21} \lambda _2 \eta _1  + \eta _2 \left( {N_{22} \lambda _2  + \frac{{2G\left( {M_{\text{MW}}^4  + M_{\text{M31}}^4 } \right)}}{{M_{\text{M31}}^3 \left| {\eta _2^{\text{M31}} } \right|^3 }}} \right)}  \\
   {\eta _3 \left( {\lambda _3  - \frac{{G\left( {M_{\text{MW}}^4  + M_{\text{M31}}^4 } \right)}}{{M_{\text{M31}}^3 \left| {\eta _2^{\text{M31}} } \right|^3 }}} \right)}  \\
\end{array}} \right)
\]

In the same way we can derive the term due to the apparent force.
Switching to the revolving system $S_2 $ we apply another rotation
expressed by  $\bm{N}{\bm{\Omega  \times }}\left( {\bm{N}{\bm{\Omega
\times }}\bm{N}{\bm{\eta }}} \right) = \bm{N}\left( {{\bm{\Omega
\times }}\left( {{\bm{\Omega  \times \eta }}} \right)} \right)$
where ${\bm{\Omega }} = {\bm{\Omega }}\left(
{{\bm{\Omega }}\left( {S_1 } \right),{\bm{\Omega }}\left(
{S_2 } \right)} \right)$ is the angular velocity of the system $S_2
$ as seen from $S_0 $. The resulting centrifugal term is
\[
{\bm{\Omega }} \times \left( {{\bm{\Omega }} \times {\bm{\eta }}}
\right) = \left( {\begin{array}{*{20}c}
   { - \left( {N_{11} \eta _1  + N_{12} \eta _2 } \right)\Omega _3^2 }  \\
   { - \left( {N_{21} \eta _1  + N_{22} \eta _2 } \right)\Omega _3^2 }  \\
   0  \\
\end{array}} \right).
\]
The associated equilibrium energy state is  given by the solution of
the systems
\begin{equation}\label{E30}
    \left\{ \begin{array}{l}
  - N_{12} \eta _2 \left( {\lambda _1  + \Omega _3^2 } \right) + \eta _1 \left( {\gamma  - N_{11} \lambda _1  - N_{11} \Omega _3^2 } \right) = 0 \\
  - 2\gamma \eta _2  - \left( {N_{21} \eta _1  + N_{22} \eta _2 } \right)\left( {\lambda _2  + \Omega _3^2 } \right) = 0 \\
 \left( {\gamma  - \lambda _3 } \right)\eta _3  = 0 \\
 \end{array} \right.
\end{equation}
where
\begin{equation}\label{E31}
\gamma  \equiv \frac{{G\left( {M_{\text{MW}}^4  + M_{\text{M31}}^4 }
\right)}}{{M_{\text{M31}}^3 \left| {\eta _2^{\text{M31}} } \right|^3 }} > 0\,\forall
t.
\end{equation}

\textit{The system \eqref{E30} is the result we are looking for. It
is evident from this system of equations that the plane $\eta _3 =
0$ is the equilibrium plane for the dynamical evolution of the
gravitational system.}

The above result is fully adequate for our purposes because the total
potential is separated into the  radial and vertical components.
This is a standard consequence of the linear approximation obtained by truncating
the Taylor expansion of the potential at the second order. Equivalently one could use a generating function satisfying the Stakel theorem for separability in a Hamilton-Jacobi equation for the above system (e.g
\cite{1998ApL&C..35..461B}).

As far as  the temporal  evolution is concerned ($t \in \left] {t_{\text{inf} }, t_0} \right]$), we have $\eta _3
\left( t \right) = 0$ which, translated into our spatial resolution,
simply  means a physical spatial resolution $\left| {\eta _3 } \right| <100 \, \rm kpc$.
Furthermore,  one can never have $\left( {\gamma  - \lambda _3 }
\right) = 0 $ because this would imply  that $\exists\, t:\lambda _3 \left( t
\right) = \gamma \left( t \right) $, whereas according to
definition \eqref{E31} we have $\gamma \left( t \right) > 0\,\forall
t$, and finally  $\lambda _3  < 0\,\forall t_{\text{lb}} < 9$ Gyr as shown by
Fig. \ref{Quadrupolo_GL_autovalori_evoluzione}. These last two conditions are \textit{clearly inconsistent} leading to a contradiction that concludes our proof as required.

\textbf{ \textit{ Therefore, the  major conclusion of this
demonstration is that the only possible solution is the following
one: the statistical minimization of PC07 is compatible with a
planar distribution of the dwarfs and it is not a mere coincidence.}}
This result completes the missing interpretation of the result
already included in PC07.

\subsection{Stability of the equilibrium configuration}
Finally,  we examine the stability of the equilibrium
plane that we have found by solving Eqn. \eqref{E30}. We take the energy of
Eqn. \eqref{E25} written for the system $S_2$
\[
\frac{{\partial E}} {{\partial {\bm{\eta }}}} =
m{\bm{N}}\left[ {{\bm{\Omega }} \times \left(
{{\bm{\Omega }} \times {\bm{\eta }}} \right)} \right] -
m{\bm{O}}^T {\bm{T{\rm O}N\eta }}
\]
and calculate the derivative
    \[
\frac{{\partial ^2 E}} {{\partial {\bm{\eta }}\partial
{\bm{\eta }}}} = m\frac{\partial } {{\partial \eta _\text{g} }}\left\{
{N_{\text{ij}} \varepsilon _{\text{jkl}} \Omega _\text{k} \varepsilon _{\text{lmn}} \Omega _\text{m}
\eta _\text{n}  - \left[ {{\bm{\Gamma N}}} \right]_{\text{ij}} \eta _\text{j} }
\right\},
\]
where $\varepsilon _{\text{ijk}}$ is the Levi-Civita symbol, summation over the
repeated index is assumed and we define $\left[ {\bm{\Gamma N}}
\right]_{\text{ij}}  \equiv \left[ {\bm{O}^T \bm{TON}} \right]_{\text{ij}}$.
Indicating with ${\bm{S}} = {\bm{ON}}$ the composition of the two
rotation matrices ${\bm{O}}$ from $S_0 $ to $S_1 $  and ${\bm{N}}$
from $S_1 $ to $S_2 $, after some algebraic simplifications we obtain
a more compact form

\begin{equation}\label{E32}
\frac{{\partial ^2 E}}{{\partial {\bm{\eta }}\partial {\bm{\eta }}}}
=  - m{\bm{N}}\left\{ {{\bm{\Theta  + S}}^T {\bm{{\rm T}S}}}
\right\}
\end{equation}
where we have defined another matrix
    \[
\begin{gathered}
  \Theta _{\text{ij}}  \equiv \left\| {\bm{\Omega }} \right\|^2 \delta _{\text{ij}}  - \Omega _\text{i} \Omega _\text{j}  =  \hfill \\
   = \left( {\begin{array}{*{20}c}
   {\Omega _2^2  + \Omega _3^2 } & { - \Omega _1 \Omega _2 } & { - \Omega _1 \Omega _3 }  \\
   { - \Omega _2 \Omega _1 } & {\Omega _1^2  + \Omega _3^2 } & { - \Omega _2 \Omega _3 }  \\
   { - \Omega _3 \Omega _1 } & { - \Omega _3 \Omega _2 } & {\Omega _1^2  + \Omega _2^2 }  \\

 \end{array} } \right) \hfill \\
\end{gathered}
\]
This matrix has no inverse, determinant $\left| \Theta \right| = 0$,
and  trace $Tr\left( \Theta  \right) = 2\left\| {\bm{\Omega }}
\right\|^2 $.

At this stage the usual procedure would be to solve for the eigen-system
\eqref{E32}. However, we can  avoid this complication by
noticing that in the linear approximation the  vertical potential
decouples from the radial one. In other words, applying the full
procedure  we would obtain three eigenvalues $\left\{ {\alpha _1
\left( t \right),\alpha _2 \left( t \right),\gamma  - \lambda _3
\left( t \right)} \right\}$, where $\alpha _{1} \left( t \right)$
and $\alpha _{2} \left( t \right)$ are two complicated functions of
the time, whereas the last eigenvalue has to be exactly the most
negative eigenvalue of the external tidal field for which we have
already calculated the time dependence as shown in Fig.
\ref{Quadrupolo_GL_autovalori_evoluzione}. As long as the eigenvalue
remains negative,  the relation $\gamma \left( t \right) - \lambda
_3 \left( t \right) > 0 \Leftrightarrow \gamma \left( t \right) >
\lambda _3 \left( t \right)$ holds because $\gamma
> 0$ by definition (equation \eqref{E30}) whereas $\lambda _3 < 0$ as required in Fig.
\ref{Quadrupolo_GL_autovalori_evoluzione}. \textit{\textbf{Therefore
the plane $\pi$  is stable.}}

\subsection{Equations of motion and force balance}
Finally, we can get a much deeper insight for the physical reasons
of the existence of the plane $\pi$ by analyzing
 the equation of motion in $S_2 $. If the plane $\pi $
 is a stable configuration of the spatial distribution of dwarf galaxies,
 it is important to isolate the force acting
on it. The equations of motion in $S_2$ are, from e.g. Eqn. \eqref{E24},
\[
\begin{gathered}
  {\bm{\ddot \eta }} = {\bm{N}}^T \left( {\left\langle {{\bm{T}}_{\text{ext}} \left( {\bm{R}} \right) + {\bm{T}}_{\text{MW}} \left( {\bm{R}} \right) + {\bm{T}}_{\text{M31}} \left( {\bm{R}} \right),{\bm{N\eta }}} \right\rangle } \right) -  \\
  2{\bm{\Omega }} \times {\bm{\dot \eta }} - {\bm{\dot \Omega }} \times {\bm{\eta }} - {\bm{\Omega }} \times \left( {{\bm{\Omega }} \times {\bm{\eta }}} \right) \\
\end{gathered}
\]
where for the different elements of the tidal tensor we can now
write
$$\begin{array}{l}
 \bm{N}^T \bm{T}_{\text{M31}} \bm{N}{\bm{\eta }} =  \\
  = \left( {\begin{array}{*{20}c}
   { - \frac{{GM_{\text{M31}} \eta _1^{\text{M31}} }}{{\left\| {{\bm{\eta }}_{\text{M31}} } \right\|^3 }}} & 0 & 0  \\
   0 & {\frac{{2GM_{\text{M31}} \eta _2^{\text{M31}} }}{{\left\| {{\bm{\eta }}_{\text{M31}} } \right\|^3 }}} & 0  \\
   0 & 0 & { - \frac{{GM_{\text{M31}} \eta _3^{\text{M31}} }}{{\left\| {{\bm{\eta }}_{\text{M31}} } \right\|^3 }}}  \\
\end{array}} \right), \\
 \end{array}$$
an analogous equation for MW, and
    \[
\bm{N}^T \bm{T}_{\text{ext}} \bm{N} = \left( {\begin{array}{*{20}c}
   {T_{11} } & {T_{12} } & 0  \\
   {T_{12} } & {T_{22} } & 0  \\
   0 & 0 & {\lambda _3 }  \\

 \end{array} } \right).
\]
The equilibrium in the $\pi$ plane is given by
\[
\left\{ \begin{array}{l}
 \left( {T_{11}  - \frac{{GM_{\text{M31}} }}{{\left\| {{\bm{\eta }}_{\text{M31}} } \right\|^3 }} - \frac{{GM_{\text{MW}} }}{{\left\| {{\bm{\eta }}_{\text{MW}} } \right\|^3 }}} \right)\eta _1  +  \\
    T_{12} \eta _2  + \eta _1 \Omega _3^2  - 2\Omega _3 \dot \eta _2 + \dot\Omega_3\eta_2 = 0 \\
 T_{12} \eta _1  + \left( {T_{22}  + \frac{{2GM_{\text{M31}} }}{{\left\| {{\bm{\eta }}_{\text{M31}} } \right\|^3 }} + \frac{{2GM_{\text{MW}} }}{{\left\| {{\bm{\eta }}_{\text{MW}} } \right\|^3 }}} \right)\eta _2  +  \\
    \eta _2 \Omega _3^2  + 2\Omega _3 \dot \eta _1  -\dot\Omega_3\eta_1 = 0 \\
 \eta _3 \left( { - \frac{{GM_{\text{M31}} }}{{\left\| {{\bm{\eta }}_{\text{M31}} } \right\|^3 }} - \frac{{GM_{\text{MW}} }}{{\left\| {{\bm{\eta }}_{\text{MW}} } \right\|^3 }} + \lambda _3 } \right) = 0 \\
 \end{array} \right.
\]
which sheds light on what is happening in reality\footnote{We did not exploit here the Eqn. \eqref{Equetas} previously necessary for the energy analysis, in favor of a clearer and easier physical interpretation of the terms in the equation.}.

Along  the direction orthogonal to the plane $\pi $, three  forces
are present.  One is due to the MW, $ - \frac{{GM_{\text{MW}} }}{{\left\|
{{\bm{\eta }}_{\text{MW}} } \right\|^3 }}$, another due to M31, $ -
\frac{{GM_{\text{M31}} }}{{\left\| {{\bm{\eta }}_{\text{M31}} } \right\|^3 }}$,
and the third one due to the external field $ + \lambda _3 < 0$. They
all sum together to push any dwarf which tends to escape from the
position of equilibrium to return back in the plane.
\textit{\textbf{Therefore they tend to flatten the whole system}}.
This tendency of flattening the distribution of dwarf galaxies, and
claiming the stability of the plane $\pi$, can be considered valid
in the limit of the linear approximation, i.e. roughly for 150 kpc
above and below the plane and for a period of time of roughly 9 Gyr,
thus being partially able to imprint the initial proper motions of
the dwarf galaxies taken into consideration for the Local Group.

Along the directions parallel to the plane, the situation is  more
complicated and described by relations like
\[
\left( {T_{11}  - \frac{{GM_{\text{M31}} }}{{\left\| {{\bm{\eta }}_{\text{M31}} }
\right\|^3 }} - \frac{{GM_{\text{MW}} }} {{\left\| {{\bm{\eta }}_{\text{MW}} }
\right\|^3 }}} \right)\eta _1+
\]
\[
 T_{12} \eta _2  + \eta _1 \Omega
_3^2  - 2\Omega _3 \dot \eta _2 + \dot\Omega_3\eta_2  = 0
\]
The first two terms in the sum on the left hand side are due to the tidal
field that has to balance the third term due to the centrifugal
force, the last term is due to the Coriolis effect and the extra term deriving from a non-uniform rotation of the system $S_2$. The same
kind of relation holds for the other coordinates.

\section{Summary, conclusions and consideration on the limits of the approach used}\label{conclusion}

The spatial distribution of the galaxies in the LG is the footprint of
its formation mechanism, the internal gravitational interactions
among the galaxies, and the gravitational action of external massive
galaxies or galaxy groups on the LG members.

In this paper, we have thoroughly addressed the whole subject
focusing the attention on the role played by the tidal force field
exerted by external galaxies or galaxy groups on the dwarf galaxies
of the LG, excluding those that are clearly under the dominant
gravitational effects of the HGs, in shaping the large-scale
distribution of the LG galaxies.

The idea stands on the well known effect of tidal interactions,
which can be expressed as a function of the gradient in the
gravitational force. While the gravitational force never changes
sign, its gradient can do so. Moreover, while the gravitational
force field  at any distance from the center of mass of a system
depends only on the inner distribution of matter, the tidal force
field does not; it is indeed the result of both internal and
external distributions of matter. The tidal force acting on a body
moving along a certain direction will pull it away from the origin
of the reference frame and, at the same time, push it along
directions perpendicular to the motion toward the origin of the
reference frame.  Therefore \textit{a system subjected to tidal
interactions can undergo not only tidal stripping but also tidal
compression. In other words, the space distribution of galaxies
undergoing tidal interactions tends to become flat}.

The results of this study can be summarized as follows:

\begin{itemize}
\item  The tidal forces
can be the cause of the planar distribution of these dwarf galaxies.
We analytically obtain the same numerical results of
\citet[][]{PC07}. In fact, we prove that a planar distribution of
all dwarf galaxies, excluding those tightly bounded to a HG, is compatible
with the presence of an external force field.

\item The planar geometrical distribution found
by \citet[][]{PC07} was  not known to relate to the most negative
eigenvalue (and associated eigenvector) of the tidal tensor. In that sense
this previous work was partially incomplete. Here we have gone deeper into this
issue following the original idea of \citet{1989MNRAS.240..195R} to
check whether the planar distribution is a mere coincidence or the
consequence of fundamental laws of mechanics. To address this, we first
check, using different arguments, the coincidence between the
direction given by the vector orthogonal to the geometrical plane
and that corresponding to the eigenvector with the most negative
eigenvalue. Second, we
analyze the energy of the orbital motion of the LG galaxies and find
that the minimum energy corresponds to a planar  distribution which
is exactly the geometrical plane $\pi$.
Therefore the planar distribution is the consequence  of the long
time-scale influence of the tidal forces exerted by massive galaxies
or galaxy groups external to the LG. Finally, we demonstrate that
this situation has been stable over the past 9 Gyr.

\item The equilibrium and stability of the plane is a consequence of the
minimum in the Action and of the orbits that come from this minimum. Although \citet[][]{PC07} have carefully investigated the
nature of the Minimum Action (whether local or absolute), the
uncertainty affecting the orbits derived from the Action
minimization could lead to an uncertainty in the energy analysis, the stability of the plane and the compression effect in turn here evidenced. This problem is still unsolved and
cannot be resolved at the present time because better observational
data would be required (proper motions, velocities, distance moduli,
absolute positions, masses, etc.). Hence, the analytical approach
presented here still requires further support from independent
arguments. Along this line of thought is the study of
\citet{2008ApJ...678..187V} who find  compatibility between their
results and what in \citet[][]{PC07}.

The completeness of the sample of external galaxies listed in Table \ref{Tabella01} is another factor influencing results developed here as well as in \citet{PC07}. In the previous paper special care was taken to confirm the results based on the minimization of the Action with an extended catalog from \citet{2001ApJ...554..104P}. Moreover, we can confirm an excellent concordance between the actual direction of the quadrupole tensor eigenvectors in \citet{1989MNRAS.240..195R} and that  derived here with the independent catalog compilation of \citet{2001ApJ...554..104P}. Nevertheless the continuous discoveries of new LG members suggest that the actual census of the LG galaxies cannot yet be considered complete (e.g. \citet{2008MNRAS.386.2221L}).

\item The Minimum Action together with the study of the equilibrium
through the first derivative of the energy of the system is a
method that can be used to constrain the energy of dwarf galaxies with
unknown proper motions. The missing proper motion prevents us from having
the complete energy of any given dwarf, but the minimization of the
action, together with the study of the tidal tensor, permits us
to handle the derivative of the energy in this particular situation.

\item We can find another example of a flat structure in the super-galactic plane, a slab of roughly $ \cong 25$ Mpc in thickness and of a diameter greater than 110 Mpc \citep{2000MNRAS.312..166L}, which was  already proposed by de Vaucouleurs in 1953 \citep{1953AJ.....58...30D}. All the groups of galaxies used here (Table \ref{Tabella01}) lie within this plane. With respect to this super-galactic plane the normal of the plane $\pi$ has direction $(SGL,SGB)=(95^\circ, 69^\circ)$. This plane has been determined requiring that the vector joining the MW and M31 explicitly belong to $\pi$.
If you attempt to find a best fit determination for the plane, say $\tilde{\pi}$, of the overall sample of the LG galaxies, without forcing the MW and M31 to belong to such a plane, we have, from PC07, $(l,b)=(46^\circ,29^\circ)$, i.e. $(SGL,SGB)=(93^\circ, 67^\circ)$, for the normal to the plane $\tilde{\pi}$. Thus, this is even closer to the direction of the vector associated with the most negative eigenvalue of the tidal tensor but slightly more distant from the direction of the super-galactic North Pole.
The proximity of the direction of the normal to $\pi$ and the super-galactic North Pole cannot yet be claimed as a significant result without further investigation. However, it is also not yet possible to study this problem fully, given the incompleteness of the catalogue of the nearby galaxies (see \citet{2004AJ....127.2031K}). The mass estimation together with the distances for the systems involved are still the major source of the errors we considered. Finally, in none of the external groups taken into account in Table \ref{Tabella01} can we claim the existence of a similar flat distribution of dwarf galaxies that could be a first indication of a common external effect acting on all these groups.
It seems that the local dwarf galaxies are primarily influenced the external potential of nearby groups, not by the larger and more distant mass accumulation responsible for the super-galactic plane.

\item The issue of the spatial distribution of nearby satellite
galaxies bound to their HG remains to be addressed. These dwarf galaxies have
orbits whose typical distance lies inside the dark matter halo of the hosting galaxy.
Many N-body simulations have shown that  galaxies in close proximity
of a HG are subjected to strong collisional interactions. The
satellite dwarf galaxies may suffer kick-off, interactions and may be absorbed
into the halo of  the HG (MW and M31 in our case),  and may abruptly
change the phase-space density distribution function. In other
words, the large-scale description of the gravitational interaction
adopted here and in \citet[][]{PC07} does not work at the short distance
scales of the HG-dwarf satellite interactions. This will be the
subject of a forthcoming study, in which simulations of the
interaction between a HG and its satellites will be investigated in
the framework of strong collisional dynamics along the line of work
already initiated by \citet{PaChiCar03} and \citet{Pasettetal}.

\end{itemize}

\begin{acknowledgements}
We thank the anonymous referee for useful comments and substantial improvement in the first release of the paper. We thank E. Grebel for stimulating discussions and a critical reading of the paper. We thank S. Jin for careful reading of this manuscript.
C.C. is pleased to acknowledge the hospitality and stimulating environment
provided by the Max-Planck Institut f\"ur Astrophysik in Garching during
his visits in July 2007 and February 2008. This study has been financed  by the
University of Padua by means of a postdoc fellowship to S.P. and by EARA funds.
\end{acknowledgements}

\bibliographystyle{apj}
\bibliography{BiblioArt}

\end{document}